\documentclass{iaus}
\usepackage{graphicx}

\title[IAUS 254.~~The Galactic disk in the cosmological context] {The
influence of star clusters on galactic disks: new insights on
star-formation in galaxies}

\author[Pavel Kroupa]   
{Pavel Kroupa}

\affiliation{Argelander-Institut f\"ur Astronomie, University of Bonn,
Auf dem H\"ugel 71, D-53121 Bonn, Germany\\email: {\tt pavel@astro.uni-bonn.de}}

\volume{254}  
\pagerange{119--126}
\jname{The Galactic disk in the cosmological context}
\editors{J. Andersen, J. Bland-Hawthorn \& B. Nordstrom}
\begin{document}

\maketitle

\begin{abstract}
Stars form in embedded star clusters which play a key role in
determining the properties of a galaxy's stellar population. A large
fraction of newly born massive stars are shot out from dynamically
unstable embedded-cluster cores spreading them to large distances
before they explode.  Embedded clusters blow out their gas once the
feedback energy from the new stellar population overcomes its binding
energy, leading to cluster expansion and in many cases dissolution
into the galaxy. Galactic disks may be thickened by such processes,
and some thick disks may be the result of an early epoch of vigorous
star-formation.  Binary stellar systems are disrupted in clusters
leading to a lower fraction of binaries in the field, while long-lived
clusters harden degenerate-stellar binaries such that the SNIa rate
may increase by orders of magnitude in those galaxies that were able
to form long-lived clusters.  The stellar initial mass function of the
whole galaxy must be computed by adding the IMFs in the individual
clusters.  The resulting integrated galactic initial mass function
(IGIMF) is top-light for SFRs$\,<10\,M_\odot$/yr, and its slope and,
more importantly, its upper stellar mass limit depend on the
star-formation rate (SFR), explaining naturally the mass--metallicity
relation of galaxies. Based on the IGIMF theory, the re-calibrated
H$\alpha$-luminosity--SFR relation implies dwarf irregular galaxies to
have the same gas-depletion time-scale as major disk galaxies,
implying a major change of our concept of dwarf-galaxy evolution.  A
galaxy transforms about 0.3~per cent of its neutral gas mass every
10~Myr into stars.  The IGIMF-theory also naturally leads to the
observed radial H$\alpha$ cutoff in disk galaxies without a radial
star-formation cutoff. It emerges that the thorough understanding of
the physics and distribution of star clusters may be leading to a
major paradigm shift in our understanding of galaxy evolution.

\end{abstract}

\firstsection 
\section{Introduction}
\label{pk_sec:intro}
Observations of star-formation in the solar vicinity suggest that the
majority of stars form in embedded clusters of which only a small
fraction survive to become open clusters (\cite[Lada \& Lada
2003]{LadaLada03}). A similar finding has emerged for extragalactic
systems (\cite[Larsen 2002a]{Larsen02a}, \cite[Larsen
2002b]{Larsen04a}). And in those galaxies where the star-formation
rate is high such that very massive young clusters appear,
\cite{Larsen02b} and \cite{Larsen04b} argue that there is also no
physically meaningful distinction between a ``globular-cluster''
star-formation mode and a ``galactic-disk'' star-formation mode; a
continuum of young-cluster masses is evident.  Observations of the
embedded cluster mass function (ECMF) show it to be a power-law, with
$\beta\approx 2$ (the Salpeter index $\beta=2.35$ is suggested by some
studies, \cite[Larsen 2002a]{Larsen02a},\cite[Weidner et
al. 2004]{WKL04}), so that the physical processes related to star
clusters that affect galaxies on a global scale can be calculated by
integrating over the ECMF.

The problem of clustered star formation in the context of galaxies can
be split into two parts: on the one hand, the physical processes of
cluster birth and stellar dynamics within the environment posed by a
host galaxy need to be understood, and on the other how these
propagate through to galaxy scales requires illumination. 

In the following a few particular problems are addressed in turn,
rather than providing a comprehensive review of star clusters in
galactic disks.  From this compilation of problems it can be concluded
that a rather remarkable amount of galactic astrophysics follows from
relatively simple ideas. The notation used here follows that applied
in the Cambody lectures by \cite{Kroupa08}.

\section{Cluster cores as OB-star ejection engines}
\label{pk_sec:obej}
It is still not quite settled whether star clusters form mass
segregated, but if not then the time-scale for the massive stars
($m>5\,M_\odot$) to sink towards their centers can be roughly
estimated from the equipartition time-scale, $t_{\rm ms}\approx(m_{\rm
av}/m_{\rm massive})\,t_{\rm relax}$, where $m_{\rm massive}, m_{\rm
av}$ are the masses of massive and average stars, respectively, and
$t_{\rm relax}$ is the two-body relaxation time. For embedded clusters
$t_{\rm ms}$ can be very short, of the order of 0.1~Myr (e.g. the
Orion Nebula Cluster, ONC), and so determining whether mass
segregation is established by birth is a very hard observational
problem. 

Irrespective of whether the massive stars from in the cluster centre
or not, once they are there they form a dynamically unstable cluster
core which is depleted in low-mass stars that have been pushed out of
the core region.  The core decays by ejecting massive stars on a
time-scale $t_{\rm decay}\approx N_{\rm m}\times t_{\rm core, cross}$,
where $N_{\rm m}$ is the number of massive stars in the core and
$t_{\rm core, cross}$ is the core-crossing time. The core-crossing
time, $t_{\rm core, cross}=2\,r_{\rm core}/\sigma_{\rm core}$, can be
estimated from the core radius, $r_{\rm core}$, and the velocity
dispersion of the massive stars in the core, $\sigma_{\rm core}$,
whereby care must be taken to remove binary-star motions.  For
clusters such as the ONC, $t_{\rm decay}$ can be 0.01--0.1~Myr, which
again is much shorter than its age (about 1~Myr). This suggests that
the ONC may have already shot out perhaps 70~per cent of its massive
stellar content (\cite[Pflamm-Altenburg \& Kroupa
2006]{PflammKroupa06}). This is consistent with the large observed
fraction of runaway massive stars, which has been used by \cite{CP92}
to infer the initial dynamical configuration of massive stars as being
small groups of binary-rich massive stars without the presence of many
low-mass stars.

Typical ejection velocities are 5--100~km/s, and so a large fraction
of massive stars explode~15~pc to 4~kpc away from their birth site,
assuming the first and last SN occur~3 and 40~Myr after birth,
respectively. A recent study of the distribution of massive stars is
available by \cite{SR08}, who find that 91~per cent of their sample
stars can be traced to an origin in young clusters. Studying
bow-shocks produced by OB stars traveling through the interstellar
medium, \cite{GB08} ``report the discovery of three bow shocks
produced by O-type stars ejected from the open cluster NGC~6611
(M16). One of the bow shocks is associated with the O9.5Iab star
HD165319, which was suggested as one of the best examples for isolated
Galactic high-mass star formation''.

The dynamical fact that massive stars are shot out from unstable
cluster cores has barely been incorporated in galaxy evolution models,
but is likely to have some important effects. In particular, the
existence or non-existence of isolated O-star formation would be an
important test of star-formation theories, with important implications
for the existence of the $m_{\rm max}-M_{\rm ecl}$ relation that
enters critically into the IGIMF theory of Section~\ref{pk_sec:igimf}.

\section{Binary systems}

High-resolution observations of star-forming regions have been
demonstrating that the fraction of binary systems is very high such
that most if not all stars form as binaries (\cite[Duchene
1999]{Duchene99},\cite[ Kouwenhoven et
al. 2007]{Kouwenhoven_etal_07},\cite[ Goodwin et
al. 2007]{Goodwin_etal_07}).  Star-formation mostly in triple or
quadruple systems is ruled out because their dynamical decay time is
far shorter than the age of the observed stellar populations
(typically older than 1~Myr): decayed higher-order multiple systems
would add too many single stars to the population such that the
resulting binary fraction would be too low in comparison to the
observed values (\cite[Goodwin \& Kroupa 2005]{GK05}). This sets
important boundary conditions for the star-formation process, but also
for star-cluster evolution models.

Numerical models of star-clusters must therefore begin with a binary
fraction near unity (\cite[Kroupa, Aarseth \& Hurley 2001]{KAH},\cite[
de la Fuente Marcos 1997]{Fuente97},\cite[ Ivanova et
al. 2005]{Ivanova_etal_05}) to be realistic. The evolution of a
cluster then assumes a fascinating wealth of new dynamical detail as
the binary population changes through the disruption of soft binaries
and hardening of hard binaries (\cite[Giersz \& Spurzem
2003]{GS03},\cite[ Davies et al. 2006]{Davies_etal_06},\cite[ Heggie
et al. 2006]{Heggie_etal_06},\cite[ Trenti et
al. 2007]{Trenti_etal_07}, \cite[ Portegies et
al. 2007]{Portegies_etal_07}), and the cluster responds by being
heated to expand (\cite[Meylan \& Heggie 1997]{MeylanHeggie97}), but
brief cluster cooling through early binary disruption has also been
observed (\cite[Kroupa, Petr \& McCaughrean 1999]{KPM99}).

For galactic disks the main effects of initially binary-rich clusters
are as follows:

\begin{itemize}

\item
The binary fraction decreases: one rather remarkable result is that
the observed high binary fraction among late-type stars in
star-forming regions ($f_{\rm bin}\approx 1$) becomes nicely
consistent with the observed binary fraction in the Galactic disk
($f_{\rm bin}\approx 0.5$) while at the same time the observed period-
and mass-ratio distribution functions are matched as well. A
unification between star-forming and disk populations has therewith
been achieved (\cite[Kroupa 2008]{Kroupa08}).

\item
The stellar ejection rate from clusters is increased significantly
(\cite[Kroupa 1998]{Kroupa98}, \cite[ K\"upper et al.  2008]{KKB08})
leading to many more stars with velocities larger than
10~km/s. Binary-rich clusters as stellar accelerators have not been
taken into account in galaxy-evolution models.

\item
The SNIa rate increases: As argued by \cite{Shara_Hurley_02}, the
tightening of binary orbits through stellar-dynamical encounters
(through sling-shot fly-bys) in long-lived clusters leads to
degenerate stars becoming tight binaries which is a pathway to
supernova type~Ia explosions through accretion-induced or
merger-induced detonation. Long lived (i.e. massive) open clusters and
globular clusters therewith become SNIa-production engines, and the
SNIa rate may go up by a factor of many if not by orders of magnitude
in Galaxies that are able to form such clusters. This is possible in
galaxies with a sufficiently high star-formation rate (SFR, \cite[
Larsen 2004b]{Larsen04b}): Only clusters with a life-time longer than
about 5~Gyr host an environment dense enough for a sufficiently long
time to allow sufficient hardening of the degenerate binaries. For
example, clusters of mass $10^{4.5}\,M_\odot$ have such life-times in
a Milky-Way type galaxy if they orbit at a distance of about 8~kpc and
on circular orbits (\cite[Baumgardt et al. 2008]{BdeMK08}). It follows
that galaxies with $SFR>0.1-1\,M_\odot$/yr are able to produce such
clusters (\cite[Weidner et al. 2004]{WKL04}).

The implications of this type of SFR-dependence of the SNIa rate for
chemical-enrichment have not been studied.

\end{itemize}

\section{The expulsion of residual gas and thickened galactic disks}

Because only about 30~per cent of the gas within a few~pc of a
star-cluster-forming cloud core ends up in stars and the remaining gas
is expelled, the emerging young cluster expands. If gas-expulsion is
explosive, i.e. occurs on a time-scale shorter or comparable to the
crossing-time scale of the cluster, then the cluster expands with a
velocity comparable to the velocity dispersion of the
pre-gas-expulsion embedded cluster.  A remnant cluster may survive
containing a small fraction of the birth stellar population: we have a
young open cluster with an associated expanding OB association
(\cite[Kroupa et al. 2001]{KAH}).  If gas expulsion occurs on a time
scale of a few cluster crossing times or longer, then the exposed
cluster reacts adiabatically by expanding and looses a much smaller
fraction of its birth population (\cite[Lada, Margulis \& Dearborn
1984]{LMD84}, \cite[ Baumgardt \& Kroupa 2007]{BM07}).

This naturally leads to a rapid evolution of the cluster mass function
(\cite[Kroupa \& Boily 2002]{KB02},\cite[ Parmentier et
al. 2008]{Parmentier_etal_08}), such that the exposed-cluster mass
function should appear less steep than the embedded-cluster mass
function. It may thus be that the embedded-cluster mass function has a
Salpeter power-law index ($\beta=2.35$) while the exposed star-cluster
mass function appears with $\beta\approx 2$.  Disruptive expulsion of
residual gas also implies a natural explanation for the origin of the
population~II stellar halo as stemming from low-mass clusters that
formed in association with todays globular clusters (\cite[Baumgardt
et al. 2008]{BKP08}), given that there is no special mode of globular
cluster formation but rather a continuous power-law embedded-cluster
mass function, the upper mass limit of which depends on the SFR
(\cite[Weidner et al. 2004]{WKL04}).

Another implication of residual gas expulsion is that galactic disks
get thickened because each star-cluster generation has associated with
it the unbound young stars lost from the clusters within the first
Myr. If the galaxy has a star-formation rate high enough ($SFR>{\rm
few}\,M_\odot$/yr) to allow the occurrence of massive clusters, then
the expanding population has a larger velocity dispersion. A galaxy
may thus go through episodes of disk-thickening events if it
experiences episodes of increased SFR.

A decreasing stellar velocity-dispersion--age relation towards younger
stellar ages, as observed for solar-neighbourhood stars, can be
accounted for quite naturally if the Milky Way has a somewhat falling
SFR over the past 5--10~Gyr, whereby proper account of other
disk-heating mechanisms such as scattering of stars on molecular
clouds, on spiral density waves and the bar must be taken into account
appropriately (\cite[Kroupa 2002]{Kroupa02}).

The Milky Way (MW) thick disk may also result from this process: If
the very early MW disk went into a star-burst globally such that
compact clusters formed throughout the disk with masses ranging up to
$10^5\,M_\odot$, and if these experienced a rapid phase of residual
gas expulsion such that the ``popping'' clusters lost a large fraction
of their stars which expanded into the young MW with a velocity
dispersion comparable to the pre-gas-expulsion velocity dispersion of
the embedded clusters ($30-50\,$km/s), then a thick-disk component
with today's thick-dick velocity dispersion can result.

This scenario has received empirical support from the observation of
clumpy but straight disks in chain and spiral galaxies from redshifts
$z\sim0.5$ to~5 (\cite[Elmegreen \& Elmegreen
2006]{ElmElm06}). Elmegreen \& Elmegreen write that these observations
seem ``{\it inconsistent with the prevailing model in which thick
disks form during the violent impact heating of thin disks and by
satellite debris from this mixing.}''  The authors discount the above
``popping cluster'' scenario on the basis that the observed
``clusters'' are massive clumps of sizes of a few hundred~pc, and
because the observed thickness is not likely to be carried through to
todays disk galaxies because of adiabatic contraction as a result of
thin-disk growth. The Elmegreen \& Elmegreen high-redshift thick disks
would thus become thinner as galaxies grow to the present epoch.

However, the large and massive clumps observed in chain galaxies by
Elmegreen \& Elmegreen are most certainly massive star-cluster
complexes as observed in the Antennae galaxies, for example
(\cite[Bastian et al. 2006]{Bastian_etal_06}). Within each
star-cluster complex, each individual cluster would ``pop''
(\cite[Fellhauer \& Kroupa 2005]{FK05}), preserving the essence of the
above ``disk-thickening through popping clusters'' theory. If the
young clusters are mass-segregated (e.g. \cite[Marks et
al. 2008]{Marks_etal_08}), then mostly low-mass stars would be lost
with a velocity dispersion of 30--50\,km/s. Such a population would
form a thick oblate distribution about the observed young galaxies but
would not be detectable. As the thin disks grow to their present
masses, the spheroidal component would adiabatically contract and
appear as today's thick disk.

Clearly, this scenario for the origin of thick disks does not
dependent on satellite mergers and works in any disk galaxy that
experienced a major star-burst throughout the disk.  The ``popping
cluster'' scenario is attractive because it relies on known physics
rather than invoking for example dark-matter substructures that are
merely hypothetical at the present time, but it needs more
quantitative work in order to formulate predictable quantities such as
the parameters of the velocity ellipsoid at different positions in the
Galaxy and for different stellar age-groups. Such work is overdue,
given that the data accumulated with the GAIA mission will allow very
accurate testing of this scenario.

It should be noted that this ``popping-cluster'' scenario does not
exclude some thick-disks to have originated from infalling satellite
galaxies. Indeed, infalling satellites are the only viable scenario
for producing counter-rotating disks.

\section{The time scale for the birth of a complete star-cluster population}
\label{pk_sec:clpop}

The stellar IMF describes the distribution of stars that form together
in a cluster-forming cloud core within a time-scale of a few Myr at
most. It is the statistical outcome of the physical processes that act
in the core when star formation proceeds.

What about the star-cluster initial mass function, i.e. the mass
function of embedded clusters (ECMF)? Star clusters form in regions of
a galaxy where the molecular clouds are massive and dense enough, but
it is not clear whether an ensemble of freshly formed star clusters
can be defined that represent an initial population in the sense of
the stellar IMF. In constructing a galaxy, it is useful though to have
this tool.

In this context, it is interesting to note that \cite{Egusa_etal_04}
find a characteristic time-scale of about 5~Myr for HII regions to
appear after the inter-stellar medium assembles in molecular clouds
along spiral arms in disk galaxies. Thus, a disk galaxy would be
churning out populations of co-eval (within a few~Myr) embedded star
clusters on this time-scale. \cite{Renaud_etal_08} investigate the
regions of cluster formation in interacting galaxies, and find these
to be the fully-compressive tidal regions. The time-scale the
inter-stellar medium (ISM) spends in these regions, emanating from
them as star clusters, is 10~Myr. Thus, even in massively interacting
galaxies it seems that the ISM transforms into star clusters on a
time-scale of about 10~Myr when the conditions for star formation are
given.

An initial star-cluster population, described by an ECMF, $\xi_{\rm
ecl}(M_{\rm ecl})$, forms on some time-scale $\delta t$, and the total
mass in stars thus produced is $M_{\rm tot}=\int_{M_{\rm
ecl,min}}^{M_{\rm ecl,max}}M_{\rm ecl}\,\xi_{\rm ecl}(M_{\rm ecl})$
$\,dM_{\rm ecl} = \delta t \times SFR$. Here $M_{\rm ecl}$ is the
stellar mass in the embedded cluster, and the minimum and maximum
values are $M_{\rm ecl,min}\approx 5\,M_{\odot}$ (Taurus-Auriga-type
star-formation) and $1=\int_{M_{\rm ecl,max}}^\infty \xi_{\rm
ecl}(M_{\rm ecl})\,dM_{\rm ecl}$ since there is only one most-massive
cluster assuming there to be no bound on the physically maximum
cluster mass.

This set of equations relates the current SFR and $M_{\rm ecl,max}$.
Using the Larsen-sample of star clusters younger than about 10~Myr in
star-forming galaxies (chosen independently to the above studies and
in order to minimise the effects of early cluster dissolution),
\cite{WKL04} (WKL) fit the theoretical $M_{\rm ecl,max}$ vs $SFR$
relation to the observational data. The best-fitting relation has a
power-law ECMF with $\beta=2.4$ (i.e. Salpeter), and $\delta t\approx
10$~Myr, independently of the SFR. The remarkable finding here is that
the time-scale of forming a star-cluster population is again found to
be about 10~Myr.

It therefore appears that galaxies form star-cluster populations such
that on a time-scale of about $\delta t=10$~Myr a statistically
complete representation of the ECMF is given. This corresponds to the
duty-cycle of molecular clouds (formation from the ISM to emerging
star clusters).  This constitutes an important insight, but must be
challenged by further research, as one possible criticism is that the
WKL result may be affected by the choice of ages of the star clusters
in the sample (see also \cite[Bastian 2008]{Bastian08}).

\section{The galaxy-wide IMF and new insights on galaxy evolution}
\label{pk_sec:igimf}
Essentially all of today's understanding of how galaxies evolve and
appear rests on the assumption that the stellar initial mass function
(IMF) is universal, being roughly a Salpeter-power-law (with index
$\alpha=2.35$, \cite[Salpeter 1955]{Salpeter55}) perhaps with a
flattening below $1\,M_\odot$. The ``{\it canonical IMF}'' has
$\alpha_1=1.3$ for $0.08-0.5\,M_\odot$ and $\alpha_2=2.3$ for
$0.5<m/M_\odot<1$ and $\alpha_3=\alpha_2$ for $m>1\,M_\odot$, while
the Scalo-field IMF has $\alpha_3\approx 2.7$ (\cite[Scalo
1986]{Scalo86}).

With this assumption, the mass-metallicity relation of galaxies needs
outflows to carry away metals from dwarf galaxies as otherwise the
bend-down of the metal-abundances towards lower-mass galaxies cannot
be understood (\cite[Kobayashi, Springel \& White
2007]{Kobayashi_etal_07}). At the same time, the outflows must not
remove the gas from late-type dwarf galaxies as these are gas-rich
today, with gas-to-stellar mass ratio near 0.8 (as opposed to MW-type
disk galaxies where it is about~0.2). This poses a problem for the
outflow scenario.  Also, the galaxy-wide SFR is commonly calculated
from a measured H$\alpha$ luminosity, and the widely-used SFR($L_{{\rm
H}\alpha}$) relation (\cite[Kennicutt et al. 1994]{Kennicutt_etal_94})
is linear because the number of ionising massive stars scales linearly
with the number of stars formed if the IMF is taken to be invariant.

Based on the neutral-gas masses measured by 21$\,$cm observations and
the SFRs measured with the H$\alpha$ flux, very low star formation
``efficiencies'' (i.e. extremely {\it long} gas-consumption
time-scales of many Hubble times) are deduced for dwarf galaxies,
while MW-type galaxies have relatively high efficiencies, i.e. {\it
short} gas-depletion time-scales of about $3\,$Gyr.

Only recently has it been fully realised that the assumption of an
invariant IMF for galaxies needs revision (\cite[Kroupa \& Weidner
2003]{KW03}): The IMF for a galaxy is given, mathematically, by the
summation of all IMFs in all forming star clusters. In each star
cluster the IMF is the same invariant parent distribution (the above
canonical form), except that the stellar masses are bounded above by
the available mass in the pre-cluster cloud core. This is a rather
elementary physical constraint: for example, star-forming cloud cores
of a few~$M_\odot$ as in Taurus-Auriga-like stellar clusters
containing a dozen stars, cannot form stars that weigh more than a
few~$\,M_\odot$. This issue that the summed IMF of many star clusters
cannot be the same as the IMF had already been concluded correctly by
\cite{Vanbeveren82}.

\subsection{The $m_{\rm max}-M_{\rm ecl}$ relation}

The existence or non-existence of a physical maximum stellar-mass ---
star-cluster-mass, $m_{\rm max}-M_{\rm ecl}$, relation (\cite[Weidner
\& Kroupa 2006]{WK06} and references therein), is of much importance
for the {\it IGIMF theory} described below, and can be understood in
terms of feedback termination of star-formation in a cloud core
through the radiation and winds of the most massive stars together
with a time-sequence of stellar-mass buildup such that low-mass stars
form, on average, before the massive stars are able to destroy the
cloud core.

Since this relation is rather fundamental, not only for our
understanding of how a typical star-cluster is assembled, but also for
the development of the IGIMF theory, it is worth-while to spend a few
words on the $m_{\rm max}-M_{\rm ecl}$ relation: A distribution of
stars always arises within a cloud core, given that the cores are
turbulent and thus have a distribution of density maxima (\cite[Padoan
\& Nordlund 2002]{PadoanNordlung02}, \cite[ Li, Klessen \& Mac Low
2003]{Li_etal_03}).  The stars formed add up to the stellar mass in
the core, $\Sigma_{\rm stars} = M_{\rm ecl}$. And so the true maximum
stellar mass, $m_{\rm max}$, that can form within a cluster is limited
even more strictly than $m_{\rm max}\le M_{\rm ecl}$, because $m_{\rm
max}=M_{\rm ecl}$ would be a cluster consisting of one star leaving no
room for the range of stellar masses resulting from turbulent gas
dynamics. In other words, isolated O-star formation cannot occur. 

However, \cite{MaschbergerClarke08} used observational $m_{\rm
max}-M_{\rm ecl}$ data to suggest that the existence of a physical
$m_{\rm max}-M_{\rm ecl}$ relation cannot be confirmed
empirically. But, they also note that the random IMF sampling model,
which admits isolated O~stars (\cite[Parker \& Goodwin
2007]{ParkerGoodwin07}), is inconsistent with the data. Importantly,
the data as used by \cite{MaschbergerClarke08} cannot be interpreted
in terms of a currently existing theoretical model.  They can,
however, be understood rather straightforwardly (Oh et al., in
preparation) if the $m_{\rm max}-M_{\rm ecl}$ relation exists {\it
and}~(1) if OB stars are shot out from binary-rich and mass-segregated
clusters {\it appearing} as isolated O~stars and~(2) through the rapid
dissolution of intermediate-mass clusters through explosive
gas-expulsion. This latter process leads to the most massive stars
being surrounded by the feeble remnant of the once existing embedded
cluster, and thus this $m_{\rm max}-M_{\rm ecl}$ datum would be an
outlier at too low a value of observed $M_{\rm ecl}$. Since virtually
all those data that do deviate from the relation are indeed outliers
at too small $M_{\rm ecl}$ values, this explanation would appear very
natural.

With this in view, the data that do not lie along the proposed $m_{\rm
max}-M_{\rm ecl}$ relation which are, however, used by Maschberger \&
Clarke, have been removed by \cite{WK06} because they are typically
older objects. Instead, \cite{Weidner_etal_07} used these data to
estimate the damage to star clusters done by the removal of residual
gas in order to constrain the star-formation efficiency in the birth
clusters under the assumption that these data were originally on the
relation.  This demonstrates the difficulty in interpreting the
existing data (Maschberger \& Clarke and Parker \& Goodwin {\it vs}
Weidner \& Kroupa).

In essence, the argument is concerned with the question whether a {\it
physical} $m_{\rm max}-M_{\rm ecl}$ relation exists, or whether
star-formation in a star cluster is a mere statistical affair such
that stellar masses appear stochastically without any physical
boundary conditions. The existence of a pronounced $m_{\rm max}-M_{\rm
ecl}$ relation for star clusters would imply a self-regulatory
behaviour of star-formation on cluster-forming molecular-cloud-core
scales. The outline of this physical process would be that as the
cloud core begins to contract the physical conditions within it are
probably similar to what we observe in Taurus-Auriga. As the core
contracts the density increases such that ever more massive
proto-stars can condense until their feedback energy suffices to
overcome gravitational collapse and the process halts, leaving an
exposed and probably largely unbound stellar cluster. This would be a
deterministic process in the sense that the initially available gas
mass within a given region, i.e. the pressure, would determine the
type of star cluster and the mass of its most massive stars, whereby
the IMF remains close to the canonical value (as determined by
observations).  Theoretical considerations based on this line of
thought do indeed yield a well-pronounced $m_{\rm max}-M_{\rm ecl}$
relation (\cite[Weidner et al. 2008]{Weidner_etal_08}).

A quantitative counter-argument against the non-existence of a
physical $m_{\rm max}-M_{\rm ecl}$ relation is, finally, as follows:
First of all, the observational evidence is such that in all cases of
well-resolved clusters, the IMF is always found to be consistent with
the canonical IMF. Now, assuming a purely random-sampling model such
that stellar masses can be generated from the canonical IMF and
$\Sigma_{\rm stars}=M_{\rm ecl}$, where $M_{\rm ecl}$ is a pre-defined
stellar cluster mass, there are occurrences such that $m_{\rm
max}\approx M_{\rm ecl}$ (\cite[Parker \& Goodwin
2007]{ParkerGoodwin07}). These would be the observed 4~per cent
isolated O~stars.  However, there would also be cases where the first
cluster star picked from the IMF is massive enough to destroy the
cloud core through feedback energy such that no further stars can
form. The resulting number of isolated O~stars would therefore
outnumber the observed number of isolated O~stars. 

Given the above and the existence of known physical processes that can
explain the existence of apparently isolated O~stars as being either
ejected stars (\S~\ref{pk_sec:obej}) or remnants of intermediate-mass
clusters that rapidly dissolved after residual gas expulsion
(\cite[Weidner et al. 2007]{Weidner_etal_07}), it follows that the
purely random sampling model for creating star clusters (``the
non-existence model'') probably needs to be rejected.

\subsection{The IGIMF}

By taking into account the existence of a physical $m_{\rm max}-M_{\rm
ecl}$ relation, i.e. that a cluster of stellar mass $M_{\rm ecl}$
cannot have stars with $m\ge m_{\rm max}={\rm fn}(M_{\rm ecl})$, and
then adding up all the so-constructed IMFs for a co-eval embedded
star-cluster population that is a power-law with Salpeter index $\beta
=2.35$ (see also \S~\ref{pk_sec:intro} and \S~\ref{pk_sec:clpop}), it
follows that the resulting ``integrated galactic IMF'' (the IGIMF) is
steeper above about~$1.3\,M_\odot$ than the canonical IMF
(\cite[Vanbeveren 1982]{Vanbeveren82}, \cite[ Kroupa \& Weidner
2003]{KW03}). This immediately solves the finding that the field-star
IMF is steeper than the IMF in individual clusters (\cite[Vanbeveren
1983]{Vanbeveren83}, \cite[Vanbeveren 1984]{Vanbeveren84}). Thus, the
Scalo-field-IMF, which has $\alpha_3\approx 3$, is unified with the
canonical IMF ($\alpha_3=2.3$) in a straight-forward way.

Also, since the maximum cluster mass, $M_{\rm ecl, max}$, that can
form within the time span $\delta t$ within a galaxy depends on the
SFR of the galaxy (\S~\ref{pk_sec:clpop} above), the IGIMF becomes SFR
dependent, such that galaxies with a low SFR have steeper IGIMFs
because of the $m_{\rm max}-M_{\rm ecl}$ relation (\cite[Weidner \&
Kroupa 2006]{WK06}, \cite[Pflamm-Altenburg et al. 2007]{PWK07}).  More
importantly, the maximum stellar mass, $m_{\rm m}$, forming in a
galaxy decreases with decreasing SFR.

For the MW, which has a SFR of a few$\,M_\odot$/yr, the IGIMF turns
out to have an index $\alpha_{\rm IGIMF}\approx3$ above a stellar mass
of $1.3\,M_\odot$ and $m_{\rm m}=150\,M_\odot$, while for a galaxy
with a SFR near $10^{-3}\,M_\odot$/yr, $\alpha_{\rm IGIMF}\approx 3.3$
and $m_{\rm m}\approx 20\,M_\odot$.  Such a variation of the IGIMF
index has been reported by \cite{HoverstenGlazebrook08} on the basis
of analysing a hundred-thousand star-forming galaxies in the SDSS
survey. Of relevance here is that the gamma-ray flux from decaying
$^{26}$Al yields a current SFR of about $4\,M_\odot$/yr for the MW if
the (Scalo) IGIMF slope, $\alpha_3=2.7\approx 3$, for the Milky Way
field is used (\cite[Diehl et al. 2006]{Diehl_etal_06}). This
constitutes an independent confirmation of the SFR-$\alpha_{\rm
IGIMF}$ relation for the case of the MW. But further confirmation, for
example by measuring the IGIMF in dwarf galaxies, would be essential
to test this theory.

\subsection{The mass-metallicity relation of galaxies}

The IGIMF theory described above immediately explains the
mass-metallicity relation of galaxies without the need of additional
physical processes (\cite[K\"oppen et al. 2007]{KWK07}). This is so
because low-mass galaxies have a deficit of massive stars per low-mass
star when compared to more massive galaxies -- they have top-light
IGIMFs. It follows that galaxy-wide metal production is compromised
increasingly with decreasing galaxy mass.

This does not mean that outflows and infall do not occur, but that
these processes probably play a secondary role in establishing the
metal content of galaxies.

Noteworthy is that in the currently established picture outflows need
to be invoked to remove the metals but such that the gas is not blown
out, given that late-type dwarfs are metal-poor and very gas rich. In
the IGIMF theory the metals are not produced in the first place and so
unwanted gas blow-out is not an issue. 

\subsection{Gas consumption time scales}

The IGIMF theory also implies that the H$\alpha$-luminosity--SFR
calibration in general use needs to be re-calibrated. This generally
used relation leads to the widely accepted result that dwarf galaxies
have very low star-formation ``efficiencies'', i.e. gas-consumption
time-scales longer than many Hubble times. \cite{PWK07} have
re-calibrated the H$\alpha$-luminosity--SFR relation based on the
above IGIMF theory. Here, a dwarf galaxy with a low SFR is producing
significantly fewer ionising photons due to the top-light IGIMF than a
massive galaxy. For a given measured H$\alpha$ luminosity the true SFR
would therefore be significantly higher than hitherto thought.

Without any parameter adjustments, the immediate result is that the
true SFRs of dwarf galaxies are orders of magnitude higher than
thought until now, implying neutral-gas-consumption time-scales of
about 3~Gyr for {\it all} galaxies that contain neutral gas, and that
the SFR is strictly proportional to the neutral gas mass of a galaxy,
$SFR=1/(3\,{\rm Gyr}) \,M_{\rm gas}$. Naturally, this leads to a very
major revision of our understanding of galaxy evolution and of the
galaxy-wide star-formation process. Note that near $M_{\rm
gas}=10^{9.5}\,M_\odot$ the \cite[Kennicutt et
al. (1994)]{Kennicutt_etal_94} finding, that the time-scale for gas
consumption is about 3~Gyr, remains valid.

The star-formation efficiency (the fraction of gas that turns into
stars) over a time-scale of $\delta t=10\,$Myr becomes $\epsilon=
(1/300)\,M_{\rm gas}$, i.e. every 10$\,$Myr a (not strongly
interacting) galaxy turns 0.3~per cent of its neutral gas mass into
stellar mass. According to the IGIMF theory, this would be true for
all not strongly interacting galaxies with neutral gas.

\subsection{The radial H$\alpha$ star-formation cutoff}

The IGIMF theory also immediately and naturally explains the existence
of an H$\alpha$ cutoff radius in disk galaxies. The current
understanding, based on applying a universal IMF to galaxies, is that
the radial H$\alpha$-emission cutoff comes about because beyond the
H$\alpha$-cutoff radius star formation is suppressed due to dynamical
processes. Basically, star-forming cloud cores cannot assemble beyond
a particular radius. However, using recent GALEX observations
(\cite[Boissier et al. 2007]{Boissier_etal_07}) show UV emission from
young stars to continue beyond this cutoff radius, in contradiction to
the theoretical work.

A {\it local IGIMF} description can be formulated rather
straightforwardly by connecting the local star-formation rate surface
density with the local neutral-gas surface density in a disk
galaxy. With this star-formation rate surface density, the local star
cluster density can be computed, and by summing them all up, a local
IGIMF (LIGIMF) follows. Integration of the LIGIMF over radial annuli
then yields the H$\alpha$ flux density in dependence of the radius,
and it follows quite trivially that it decays much more rapidly than
the star-formation rate density (\cite[Pflamm-Altenburg \& Kroupa
2008]{PflammKroupa08}). An interesting outcome of this work is that
the star-formation rate surface density is linearly proportional to
the gas surface density, $\Sigma_{\rm SFR} \approx 1/(3\,{\rm
Gyr})\,\Sigma_{\rm gas}^N$, with $N=1$.

\subsection{Summary: IGIMF}

The implications of all of this work would be that a paradigm shift in
galaxy evolution may be emerging such that the star-formation rate
(density) is proportional to the neutral gas mass (density).  Dwarf
galaxies differ from large disk galaxies only in terms of the level of
star formation because the latter have a larger neutral gas mass which
supports more star-formation activity. Further implications of this
are being looked into now.

\section{Conclusions}

The above shows that we have been missing important ingredients in our
understanding of the astrophysics of disk galaxies if the physics of
star clusters as the {\it fundamental galactic building blocks} is
omitted. As important examples the following have been noted:
\begin{itemize}

\item
Young cluster cores are stellar accelerators dispersing
massive stars over large distances from their birth sites. 

\item
The binary fraction and SN~Ia rate are defined by the star-cluster
population a galaxy has been able to generate over its life time.

\item
Galactic disks may thicken if they form ensembles of ``popping''
clusters.  

\item
Star-clusters in a statistically complete ensemble, such
that they represent the embedded cluster mass function, appear to form
on a time-scale of about 10~Myr. 

\item
At the same time, by realising that we must see galaxies as made up of
many (mostly dissolved) star clusters we are readily led to find a new
understanding of the mass-metallicity relation of galaxies, of the
radial H$\alpha$ cutoff and therewith the relation of neutral gas
content to the level of star formation.

\item
Following on from this, it emerges that dwarf irregular galaxies
consume their neutral gas content on the same time scale (about 3~Gyr)
as major disk galaxies, and that it is only the mass of neutral gas
that drives the macroscopic evolution of these systems in terms of the
buildup of stellar mass and metals.

\item
The last two points are a result of correctly counting massive stars
in galaxies in dependence of their SFRs: massive disk galaxies have a
higher SFR than low-mass irregular galaxies and consequently their
IGIMF is flatter and the upper mass limit of forming stars is higher.
The implications of this suggest a possibly major paradigm shift in
our knowledge of how galaxies evolve.

\end{itemize}

\vspace{5mm}

\noindent {\tt Acknowledgements:} I would like to thank the organisers
for the invitation to present this material at a most memorable
meeting in Copenhagen, and Jan Pflamm-Altenburg and Carsten Weidner
for very important contributions. This contribution I wrote in Vienna
and Canberra, and would like to thank the respective hosts (Gerhardt
Hensler, Christian Theis and Helmut Jerjen) for their kind
hospitality.

\end{document}